\documentstyle[12pt]{article}
\begin{document}
\def\beq{\begin{equation}}
\def\eeq{\end{equation}}
\def\bea{\begin{eqnarray}}
\def\eea{\end{eqnarray}}
\def\ve{\vert}
\def\vel{\left|}
\def\ver{\right|}
\def\nnb{\nonumber}
\def\ga{\left(}
\def\dr{\right)}
\def\aga{\left\{}
\def\adr{\right\}}
\def\rar{\rightarrow}
\def\nnb{\nonumber}
\def\la{\langle}
\def\ra{\rangle}
\def\lla{\left<}
\def\rra{\right>}
\def\ba{\begin{array}}
\def\ea{\end{array}}
\def\tep{$B \rar K \ell^+ \ell^-$}
\def\tepm{$B \rar K \mu^+ \mu^-$}
\def\tept{$B \rar K \tau^+ \tau^-$}
\def\ds{\displaystyle}



\newskip\humongous \humongous=0pt plus 1000pt minus 1000pt
\def\caja{\mathsurround=0pt}
\def\eqalign#1{\,\vcenter{\openup1\jot
\caja   \ialign{\strut \hfil$\displaystyle{##}$&$
\displaystyle{{}##}$\hfil\crcr#1\crcr}}\,}


\def\simlt{\stackrel{<}{{}_\sim}}
\def\simgt{\stackrel{>}{{}_\sim}}



\def\bos{\lower 0.5cm\hbox{{\vrule width 0pt height 1.2cm}}}
\def\boss{\lower 0.35cm\hbox{{\vrule width 0pt height 1.cm}}}
\def\aaa{\lower 0.cm\hbox{{\vrule width 0pt height .7cm}}}
\def\dol{\lower 0.4cm\hbox{{\vrule width 0pt height .5cm}}}


\title{ {\Large {\bf 
New physics effects to the lepton polarizations in the  
$B \rar K \ell^+ \ell^-$ decay } } }

\author{\vspace{1cm}\\
{\small T. M. Aliev \thanks
{e-mail: taliev@metu.edu.tr}\,\,,
M. K. \c{C}akmak\,\,,
A. \"{O}zpineci \thanks
{e-mail: altugoz@metu.edu.tr}\,\,,
M. Savc{\i} \thanks
{e-mail: savci@metu.edu.tr}} \\
{\small Physics Department, Middle East Technical University} \\
{\small 06531 Ankara, Turkey} }
\date{}

\begin{titlepage}
\maketitle
\thispagestyle{empty}

\begin{abstract}
Using the general,
model independent form of the effective Hamiltonian, the general 
expressions of the
longitudinal, normal and transversal polarization asymmetries for $\ell^-$
and $\ell^+$ and combinations of them for the exclusive 
$B \rar K \ell^+ \ell^-$ decay are found. The sensitivity of
lepton polarizations and their combinations on new Wilson coefficients are
studied. It is found that there exist regions of Wilson coefficients for
which the branching ratio coincides with the Standard Model result while 
the lepton polarizations differ substantially from the standard model prediction.
Hence, studying lepton polarization in these regions of new Wilson
coefficients can serve as a promising tool for establishing new physics
beyond the Standard Model.
\end{abstract}

~~~PACS numbers: 12.60.--i, 13.20.--v, 13.25.Hw
\end{titlepage}

\section{Introduction}

The two $B$ meson factories  BaBar and Belle , which have already started
operation, open an exciting new era in studying physics of $B$ mesons.
Both factories have already presented thrilling results on CP violation 
\cite{R1}. The
physics program of the B factories contain two main directions:
detailed study of CP violation in $B_d$ decays and precise measurement of
rare Flavor Changing Neutral Current (FCNC) processes. It is well known
that FCNC processes are very sensitive to the new physics beyond the
Standard Model (SM). So, main goal of the investigations undergoing at 
$B$ factories is to find
inconsistencies within the SM, in particular find indications for new
physics in the flavor and CP violating sector \cite{R2}.
New physics effects can appear
in rare $B$ meson decays in two different ways, either through new 
contributions to the Wilson coefficients existing in the SM or through
the new structures in the effective Hamiltonian which are absent in the SM.
Rare $B$ meson decays induced by $b \rar s(d) \ell^+ \ell^-$ transition has 
been extensively
studied in framework of the SM and its various extensions
\cite{R3}--\cite{R19}. One of the efficient ways in establishing new physics
beyond the SM is the measurement of the lepton polarization
\cite{R19}--\cite{R26}. All previous studies for the lepton polarization
have been limited to SM and its minimal extensions, except the works
\cite{R23,R26}. In \cite{R23} the analysis of the $\tau$ lepton polarization
for the inclusive $b \rar s \tau^+ \tau^-$ decay was presented in a model
independent way and in \cite{R26} lepton polarizations are investigated
using the most general model independent Hamiltonian for the 
$B \rar K^\ast \ell^+ \ell^-$ decay.

The aim of this work is studying lepton
polarizations in the exclusive $B \rar K \ell^+\ell^-$ 
decay using the general form of the effective Hamiltonian including
all possible form of interactions. 
Here we will study
$\mu$ and $\tau$ leptonic modes because of the
following two reasons. Firstly, the electron polarization is hard to measure 
experimentally, and secondly, it is well known that in the SM the normal
$P_N$ and transversal $P_T$ polarizations are both proportional to the
lepton mass and hence their measurements might be possible especially in the
$\tau^- \tau^+$ channel.

The work is organized as follows. In
section 2, using a general form of four--Fermi interaction  we derive the 
general expressions for the longitudinal,
transversal and normal polarizations of leptons. In section 3 we investigate
the sensitivity of the above--mentioned polarizations to the new Wilson
coefficients. At the end of this section we also present our conclusion.

\section{Calculation of lepton polarizations}

In this section we compute the lepton polarization asymmetries, using
the most general, model independent form of the effective Hamiltonian.
The effective Hamiltonian for the $b \rar s \ell^+ \ell^-$ transition 
in terms of twelve model independent four--Fermi
interactions can be written in the following form:  
\bea
\label{matel}
{\cal H}_{eff} &=& \frac{G_F\alpha}{\sqrt{2} \pi}
 V_{ts}V_{tb}^\ast
\Bigg\{ C_{SL} \, \bar s i \sigma_{\mu\nu} \frac{q^\nu}{q^2}\, L \,b  
\, \bar \ell \gamma^\mu \ell + C_{BR}\, \bar s i \sigma_{\mu\nu}
\frac{q^\nu}{q^2} \,R\, b \, \bar \ell \gamma^\mu \ell \nnb \\
&&+C_{LL}^{tot}\, \bar s_L \gamma_\mu b_L \,\bar \ell_L \gamma^\mu \ell_L +
C_{LR}^{tot} \,\bar s_L \gamma_\mu b_L \, \bar \ell_R \gamma^\mu \ell_R +  
C_{RL} \,\bar s_R \gamma_\mu b_R \,\bar \ell_L \gamma^\mu \ell_L \nnb \\
&&+C_{RR} \,\bar s_R \gamma_\mu b_R \, \bar \ell_R \gamma^\mu \ell_R +
C_{LRLR} \, \bar s_L b_R \,\bar \ell_L \ell_R +
C_{RLLR} \,\bar s_R b_L \,\bar \ell_L \ell_R \\
&&+C_{LRRL} \,\bar s_L b_R \,\bar \ell_R \ell_L +
C_{RLRL} \,\bar s_R b_L \,\bar \ell_R \ell_L+
C_T\, \bar s \sigma_{\mu\nu} b \,\bar \ell \sigma^{\mu\nu}\ell \nnb \\
&&+i C_{TE}\,\epsilon^{\mu\nu\alpha\beta} \bar s \sigma_{\mu\nu} b \,
\bar \ell \sigma_{\alpha\beta} \ell  \Bigg\}~, \nnb
\eea
where the chiral projection operators $L$ and $R$ in (\ref{matel}) are
defined as
\bea  
L = \frac{1-\gamma_5}{2} ~,~~~~~~ R = \frac{1+\gamma_5}{2}\nnb~,
\eea  
and $C_X$ are the coefficients of the four--Fermi interactions and
$q=p_B-p_K$ is the momentum transfer.
Note that among twelve Wilson coefficients several already exist in the SM.
The coefficients $C_{SL}$ and $C_{BR}$ correspond to $-2 m_s C_7^{eff}$ 
and $-2 m_b C_7^{eff}$ in the SM, respectively. The next
four terms in Eq. (\ref{matel}) are the vector type interactions with
coefficients $C_{LL}^{tot}$, $C_{LR}^{tot}$, $C_{RL}$ and $C_{RR}$. Two of these
vector interactions containing $C_{LL}^{tot}$ and $C_{LR}^{tot}$ do exist in the SM
as well in the form $(C_9^{eff}-C_{10})$ and $(C_9^{eff}+C_{10})$.
Therefore we can say that $C_{LL}^{tot}$ and $C_{LR}^{tot}$ describe the
sum of the contributions from SM and the new physics and they can be written
as
\bea
C_{LL}^{tot} &=& C_9^{eff} - C_{10} + C_{LL}~, \nnb \\     
C_{LR}^{tot} &=& C_9^{eff} + C_{10} + C_{LR}~, \nnb
\eea
The terms with
coefficients $C_{LRLR}$, $C_{RLLR}$, $C_{LRRL}$ and $C_{RLRL}$ describe
the scalar type interactions. The last two terms with the
coefficients $C_T$ and $C_{TE}$, obviously, describe the tensor type
interactions.    

Exclusive $B \rar K \ell^+ \ell^-$ decay is described by the matrix
element of effective Hamiltonian over $B$ and $K$ meson states, which can be
parametrized in terms of form factors.
It follows from Eq. (\ref{matel})
that in order to calculate the amplitude of the $B \rar K \ell^+ \ell^-$
decay, the following matrix elements are needed 
\bea
\label{roll}
&&\lla K\vel \bar s \gamma_\mu b \ver B \rra~,\nnb \\
&&\lla K \vel \bar s i\sigma_{\mu\nu} q^\nu b \ver B \rra~, \nnb \\
&&\lla K \vel \bar s b \ver B \rra~, \nnb \\
&&\lla K \vel \bar s \sigma_{\mu\nu} b
\ver B \rra~. \nnb
\eea

These matrix elements are defined as follows:
\bea
\label{ilk}
\lla K(p_{K}) \vel \bar s \gamma_\mu b \ver B(p_B) \rra =
f_+ \Bigg[ (p_B+p_K)_\mu - \frac{m_B^2-m_K^2}{q^2} \, q_\mu \Bigg] 
+ f_0 \,\frac{m_B^2-m_K^2}{q^2} \, q_\mu~,
\eea
with $f_+(0) = f_0(0)$,
\bea
\label{ucc}
\lla K(p_{K}) \vel \bar s \sigma_{\mu\nu}
 b \ver B(p_B) \rra = -i \, \frac{f_T}{m_B+m_K}
\Big[ (p_B+p_K)_\mu q_\nu -
q_\mu (p_B+p_K)_\nu\Big]~,
\eea
The matrix elements $\lla K(p_{K}) \vel \bar s i \sigma_{\mu\nu} q^\nu b
\ver B(p_B) \rra$ and $\lla K \vel \bar s b \ver B \rra$
can be calculated by contracting both sides of Eqs. (\ref{ilk}) and 
(\ref{ucc}) with $q^\mu$, and using equation of motion we get
\bea
\label{uc}
\lla K(p_{K}) \vel \bar s b \ver B(p_B) \rra &=&
f_0 \, \frac{m_B^2-m_K^2}{m_b-m_s}~, \\
\label{iki}                   
\lla K(p_{K}) \vel \bar s i \sigma_{\mu\nu} q^\nu b \ver B(p_B) \rra &=&   
\frac{f_T}{m_B+m_K} \Big[ (p_B+p_K)_\mu q^2 -
q_\mu (m_B^2-m_K^2) \Big]~.
\eea
Taking into account Eqs. (\ref{matel}--\ref{uc}), the matrix element of the 
$B \rar K \ell^+ \ell^-$ decay can be written as 
\bea
\label{had}
{\cal M}(B\rightarrow K \ell^{+}\ell^{-}) &=&
\frac{G_F \alpha}{4 \sqrt{2} \pi} V_{tb} V_{ts}^\ast
\Bigg\{
\bar \ell \gamma^\mu \ell \, \Big[
A (p_B+p_K)_\mu + B q_\mu \Big] \nnb \\ 
&+& \bar \ell \gamma^\mu \gamma_5 \ell \, \Big[
C (p_B+p_K)_\mu  + D q_\mu \Big]
+\bar \ell \ell \,Q
+ \bar \ell \gamma_5 \ell \, N \nnb \\
&+& 4 \bar \ell \sigma^{\mu\nu}  \ell\, (- i G ) 
\Big[ (p_B+p_K)_\mu q_\nu - (p_B+p_K)_\nu q_\mu
\Big] \nnb \\
&+& 4 \bar \ell \sigma^{\alpha\beta}  \ell \,
\epsilon_{\mu\nu\alpha\beta} \, H
\Big[ (p_B+p_K)_\mu q_\nu - (p_B+p_K)_\nu q_\mu \Big] 
\Bigg\}~.
\eea

The auxiliary functions above are defined as
\bea
\label{as}
A &=& (C_{LL}^{tot} + C_{LR}^{tot} + C_{RL} + C_{RR})\, f_+ +
2 (C_{BR}+C_{SL}) \,\frac{f_T}{m_B+m_{K}} ~, \nnb \\
B &=& (C_{LL}^{tot} + C_{LR}^{tot}+ C_{RL} + C_{RR}) \, f_- -
2 (C_{BR}+C_{SL})\,\frac{f_T}{(m_B+m_{K})q^2}\, (m_B^2-m_K^2) ~, \nnb \\
C &=& (C_{LR}^{tot} + C_{RR} - C_{LL}^{tot} - C_{RL})\, f_+ ~,\nnb \\
D &=& (C_{LR}^{tot} + C_{RR} - C_{LL}^{tot} - C_{RL})\, f_- ~, \\
Q &=& f_0 \, \frac{m_B^2-m_K^2}{m_b-m_s}\,
(C_{LRLR} + C_{RLLR}+C_{LRRL} + C_{RLRL})~,\nnb \\
N &=& f_0 \, \frac{m_B^2-m_K^2}{m_b-m_s}\,
(C_{LRLR} + C_{RLLR}-C_{LRRL} - C_{RLRL})~,\nnb \\
G &=& \frac{C_T}{m_B+m_K}\, f_T~,\nnb \\
H &=& \frac{C_{TE}}{m_B+m_K}\, f_T~,\nnb
\eea
where 
\bea
f_- = (f_0-f_+) \frac{m_B^2-m_K^2}{q^2}~.\nnb
\eea

It follows immediately from Eq. (\ref{had}) 
that the difference from the SM 
is due to the last four terms only, namely, scalar and tensor type interactions.
Using Eq. (\ref{had}) we next calculate the final lepton
polarizations for the $B \rar K \ell^+ \ell^-$ decay. For this purpose we 
define the
following orthogonal unit vectors, $S_L^{-\mu}$ in the rest frame of
$\ell^-$ and $S_L^{+\mu}$ in the rest frame of $\ell^+$, for the
polarization of the leptons along the longitudinal ($L$), transversal ($T$)
and normal ($N$) directions:
\bea
\label{pol}
S_L^{-\mu} &\equiv& (0,\vec{e}_L^{\,-}) = 
\ga 0,\frac{\vec{p}_-}{\vel \vec{p}_- \ver} \dr~, \nnb \\
S_N^{-\mu} &\equiv& (0,\vec{e}_N^{\,-}) = 
\ga 0,\frac{\vec{p}_K \times \vec{p}_-}
{\vel \vec{p}_K \times \vec{p}_- \ver} \dr~, \nnb \\
S_T^{-\mu} &\equiv& (0,\vec{e}_T^{\,-}) = 
\ga 0, \vec{e}_N^{\,-} \times \vec{e}_L^{\,-} \dr~, \\
S_L^{+\mu} &\equiv& (0,\vec{e}_L^{\,+}) = 
\ga 0,\frac{\vec{p}_+}{\vel \vec{p}_+ \ver} \dr~, \nnb \\
S_N^{+\mu} &\equiv& (0,\vec{e}_N^{\,+}) = 
\ga 0,\frac{\vec{p}_K \times \vec{p}_+}
{\vel \vec{p}_K \times \vec{p}_+ \ver} \dr~, \nnb \\
S_T^{+\mu} &\equiv& (0,\vec{e}_T^{\,+}) = 
\ga 0, \vec{e}_N^{\,+} \times \vec{e}_L^{\,+} \dr~, \nnb
\eea
where $\vec{p}_\mp$ and $\vec{p}_K$ are the three momenta of $\ell^\mp$ and
$K$ meson in the center of mass (CM) frame of the $\ell^+ \ell^-$
system, respectively. The longitudinal unit vectors $S_L^-$ and $S_L^+$ are
boosted to CM frame of $\ell^+ \ell^-$ by Lorentz transformation,
\bea
\label{bs}
S^{-\mu}_{L,\, CM} &=& \ga \frac{\vel \vec{p}_- \ver}{m_\ell}, 
\frac{E_\ell \,\vec{p}_-}{m_\ell \vel \vec{p}_- \ver} \dr~, \nnb \\
S^{+\mu}_{L,\, CM} &=& \ga \frac{\vel \vec{p}_- \ver}{m_\ell}, 
- \frac{E_\ell \, \vec{p}_-}{m_\ell \vel \vec{p}_- \ver} \dr~,
\eea
while vectors $\vec{S}_N$ and $\vec{S}_T$ are not changed by boost.

The differential decay rate of the $B \rar K \ell^+ \ell^-$ decay for
any spin direction $\vec{n}^{\,(\mp)}$ of the $\ell^{(\mp)}$, where 
$\vec{n}^{\,(\mp)}$ is the unit vector in the $\ell^{(\mp)}$ rest frame,
can be written as
\bea
\label{ddr}
\frac{d\Gamma(\vec{n}^{(\mp)})}{dq^2} = \frac{1}{2} 
\ga \frac{d\Gamma}{dq^2}\dr_0  
\Bigg[ 1 + \Bigg( P_L^{(\mp)} \vec{e}_L^{\,(\mp)} + P_N^{(\mp)}
\vec{e}_N^{\,(\mp)} + P_T^{(\mp)} \vec{e}_T^{\,(\mp)} \Bigg) \cdot
\vec{n}^{(\mp)} \Bigg]~,
\eea
where $\ga d\Gamma/dq^2 \dr_0$ corresponds to the unpolarized differential decay
rate, and   
$P_L$, $P_N$ and $P_T$ represent the longitudinal, normal and transversal
polarizations, respectively.
The expression for the unpolarized differential decay rate in Eq. (\ref{ddr}) is

\bea
\label{unp}
\ga \frac{d \Gamma}{dq^2}\dr_0 = \frac{G_F^2 \alpha^2}{2^{14} \pi^5 m_B} 
\vel V_{tb} V_{ts}^\ast \ver^2 \lambda^{1/2}(1,r,s) v \Delta~,
\eea
where 
$\lambda(1,r,s) = 1 + r^2 + s^2 - 2 r - 2 s - 2 r s$ , $s=q^2/m_B^2$, 
$r=m_{K}^2/m_B^2$ and $v=\sqrt{1-4m_\ell^2/q^2}$ is the lepton velocity.
The explicit form of $\Delta$ is

\bea  
\label{Del}
\Delta &=& -128 \lambda m_B^4 m_\ell \,\mbox{\rm Re}(A G^\ast) + 32 m_B^2
m_\ell^2 (1-r) \,\mbox{\rm Re}(C D^\ast) + 16 m_B^2 m_\ell (1-r) 
\,\mbox{\rm Re}(C N^\ast) \nnb \\
&+& 16 m_B^2 m_\ell^2 s \,\vel D \ver^2 + 
4 m_B^2  s \,\vel N \ver^2 + 16 m_B^2 m_\ell s \,\mbox{\rm Re}(D N^\ast) 
+\frac{1024}{3} \lambda m_B^6 s v^2  \,\vel H \ver^2\nnb \\
&+&4 m_B^2 sv^2  \,\vel Q \ver^2 + \frac{4}{3}\lambda m_B^4 
s(3-v^2)\,\vel A \ver^2+\frac{256}{3} \lambda m_B^6s
(3-v^2)\,\vel G \ver^2\nnb \\
&+&\frac{4}{3} m_B^4 s \Big\{2 \lambda - (1-v^2)\Big[ 2 \lambda -
3(1-r)^2 \Big] \Big\}\,\vel C \ver^2~.\nnb
\eea

The polarizations $P_L$, $P_N$ and $P_T$ are defined as:
\bea
P_i^{(\mp)}(q^2) = \frac{\ds{\frac{d \Gamma}{dq^2}
                   (\vec{n}^{(\mp)}=\vec{e}_i^{\,(\mp)}) -
                   \frac{d \Gamma}{dq^2}
                   (\vec{n}^{(\mp)}=-\vec{e}_i^{\,(\mp)})}}
              {\ds{\frac{d \Gamma}{dq^2}
                   (\vec{n}^{(\mp)}=\vec{e}_i^{\,(\mp)}) +
                  \frac{d \Gamma}{dq^2}
                  (\vec{n}^{(\mp)}=-\vec{e}_i^{\,(\mp)})}}~, \nnb 
\eea
where $P^{(\mp)}$ represents the charged lepton $\ell^{(\mp)}$ 
polarization asymmetry for $i=L,~N,~T$, i.e., $P_L$ and $P_T$ are the
longitudinal and transversal asymmetries in the decay plane, respectively,
and $P_N$ is the normal component to both of them. With respect to the direction
of the lepton polarization, $P_L$ and $P_T$ are $P$--odd, $T$--even, while
$P_N$ is $P$--even, $T$--odd and $CP$--odd. 
calculations lead to the following results for the longitudinal,
transversal and normal polarization of the
$\ell^{(\mp)}$:
\bea
\label{plm}
\lefteqn{
P_L^{(\mp)}= \frac{4 m_B^2 v}{\Delta} \Bigg\{\pm
\frac{4}{3}\lambda m_B^2 \, \mbox{\rm Re}(A C^\ast) \mp 
\frac{64}{3} \lambda m_B^2 m_\ell \, \mbox{\rm Re}(C G^\ast)
-\frac{64}{3} \lambda m_B^2 m_\ell \, \mbox{\rm Re}(A H^\ast)} \nnb \\
&&- 4 m_\ell(1-r) \, \mbox{\rm Re}(C Q^\ast)
+ \frac{256}{3} \lambda m_B^4 s \, \mbox{\rm Re} (G H^\ast)
- 4 m_\ell s \, \mbox{\rm Re} (D Q^\ast) - 2 \, \mbox{\rm Re} (N Q^\ast)
\Bigg\}~,\\ \nnb \\
\label{ptm}
\lefteqn{
P_T^{(\mp)}= \frac{\pi m_B^3 \sqrt{s \lambda}}{\Delta} \Bigg\{
\pm \frac{4}{s} m_\ell (1-r) \,\mbox{\rm Re} (A C^\ast)
\mp \frac{64}{s} (1-r) m_\ell^2 \,\mbox{\rm Re} (C G^\ast)} \\
&&\pm 4 m_\ell \,\mbox{\rm Re} (A D^\ast)
\mp 64 m_\ell^2 \,\mbox{\rm Re} (D G^\ast)
\pm 2 \,\mbox{\rm Re}(A N^\ast) \mp 32 m_\ell\,
\mbox{\rm Re} (G N^\ast) + 2 v^2 \,\mbox{\rm Re} (C Q^\ast)
\Bigg\}~,\nnb \\ \nnb \\
\label{pnm}
\lefteqn{
P_N^{(\mp)}= \frac{m_B^3 v \sqrt{s \lambda}}{\Delta}  \Bigg\{
4  m_\ell \,\mbox{\rm Im}(C D^\ast) 
+2 \,\mbox{\rm Im} (C N^\ast)
\mp 2 \,\mbox{\rm Im} (A Q^\ast) \pm
32 m_\ell \,\mbox{\rm Im}(C G^\ast)\Bigg\} }~.
\eea

From these expressions we can make the following conclusion. Contributions
coming from the SM to $P_L^-$ and $P_L^+$ are exactly the same but with the
opposite sign. However contributions coming from new interactions  to
$P_L^-$ and $P_L^+$ can have same or opposite sign. This can be useful in
looking for new physics. 

From Eq. (\ref{ptm}) we observe that at zero lepton mass limit,
contributions coming from scalar interactions survive. Similarly terms
coming from scalar and tensor interactions survive in the massless lepton
limit for $P_L^{(\mp)}$. Therefore, experimentally measured value of 
$P_{L,T}^{(\mp)}$ for the $B \rar K \mu^+ \mu^-$ can give a very promising
hint in looking new physics beyond SM. About normal polarization we can
comment as follows. One can see from Eq. (\ref{pnm}) the difference
between $P_N^-$ and $P_N^+$ (for which SM predicts $P_N^-=-P_N^+$) is again
due to the existence of the scalar and tensor interactions. Incidentally, w
e should note that a similar situation takes place for the lepton
polarizations in the $B \rar K^\ast \ell^+ \ell^-$ decay \cite{R26}. It
follows from this discussion that a measurement of the lepton polarization
of each lepton and combined analysis of lepton and antilepton polarizations
$P_L^- + P_L^+$, $P_T^- - P_T^+$ and $P_N^- + P_N^+$ can give very useful
information to constraint or to discover new physics beyond SM, which are
all zero in the SM in the limit of massless leptons. 
Therefore if in experiments nonzero value of the above
mentioned combined lepton asymmetries were observed, this can be considered
as an discovery of the new physics beyond SM.   

\section{Numerical analysis}
First of all we introduce the values of the input parameters used 
in the present work:
$\vel V_{tb} V_{ts}^\ast \ver = 0.0385$, $\alpha^{-1}=129$,
$G_F=1.17\times 10^{-5}~GeV^{-2}$, $~\Gamma_B=4.22\times10^{-13}~GeV$,
$C_9^{eff}=4.344,~C_{10}=-4.669$. 
It is well known that the Wilson coefficient $C_9^{eff}$ receives 
short as well as long distance contributions coming from the real 
$\bar c c$ intermediate states, i.e., with the $J/\psi$ family, but in this 
work we consider only short distance contributions. Experimental data on 
${\cal B}(B \rar X_s \gamma$ 
fixes only the modulo of $C_7^{eff}$.
For this reason throughout our analysis we have considered both
possibilities, i.e., $C_7^{eff} = \mp 0.313$, where the upper sign
corresponds to the SM prediction.
 
For the values of the form factors, we have used the results of
\cite{R27} (see also \cite{R28,R29}).
The $q^2$ dependence of the form factors can be represented in terms of
three parameters as
\bea
\label{fit}
F(s) = F(0) \, exp(c_1 s + c_2 s^2 + c_3 s^3)~,
\eea
where the values of parameters $F(0)$, $c_1$, $c_2$ and $c_3$ for the
$B \rar K$ decay are listed in Table 1.

\begin{table}[h]                    
\renewcommand{\arraystretch}{1.5}                        
\addtolength{\arraycolsep}{3pt}
$$
\begin{array}{|c|ccc|}
\hline
& f_+ & f_0 & f_T \\ \hline
F(0) &
0.319 &  \phantom{-} 0.319 & 0.355 \\
c_1 &
1.465 &  \phantom{-} 0.633 & 1.478\\
c_2 &
0.372 & -0.095 & 0.373\\
c_3 &
0.782 &  \phantom{-} 0.591 & 0.700\\
\hline
\end{array}   
$$
\caption{Central values of the parameters for the parametrization
(\ref{fit}) of the $B \rar K$ decay form factors.}
\renewcommand{\arraystretch}{1}
\addtolength{\arraycolsep}{-3pt}
\end{table}       
From the expressions of the lepton
polarizations we see that they all depend on $q^2$ and the new Wilson coefficients.
It may be experimentally difficult to study the dependence of the
the polarizations of each lepton on both quantities. 
Therefore we eliminate the dependence of
the lepton polarizations on $q^2$, by
performing integration over $q^2$ in the allowed kinematical region, so that
the lepton polarizations are averaged. The averaged lepton polarizations are
defined as
\bea
\label{av}
\lla P_i \rra = \frac{\ds \int_{4 m_\ell^2}^{(m_b-m_{K})^2}
P_i \frac{d{\cal B}}{dq^2} dq^2}
{\ds \int_{4 m_\ell^2}^{(m_b-m_{K})^2}
 \frac{d{\cal B}}{dq^2} dq^2}~.  
\eea

We present our results in a series figures. Note that in all figures we
presented the value of $C_7^{eff}$ is chosen to have its SM value, i.e., 
$C_7^{eff}=-0.313$. Figs. (1) and (2) depict the
dependence of the averaged longitudinal polarization $\lla P_L^- \rra$ of
$\ell^-$ and the combination $\lla P_L^- + P_L^+ \rra$ on new Wilson coefficients, 
at $C_7^{eff}=-0.313$ for $B \rar K \mu^+ \mu^-$ decay. From these figures we
see that $\lla P_L^- \rra$ is sensitive to the existence of all new
interactions except to vector and scalar interactions with coefficients
$C_{LL}$, $C_{RL}$ and $C_{RLLR}$, $C_{LRLR}$, respectively,
while the combined average $\lla P_L^- + P_L^+ \rra$ is sensitive
to scalar type interactions only. It is interesting that contributions from
$C_{RLLR}$, $C_{LRLR}$ ($C_{LRRL}$, $C_{RLRL}$) to the combined asymmetry is
always negative (positive). Therefore determination of the sign of $\lla
P_L^- + P_L^+ \rra$ can be useful in discriminating the type of the
interaction.
From Fig. (2) we see that 
$\lla P_L^- + P_L^+ \rra=0$ at $C_X=0$, which confirms the SM result as
expected.
For the other choice of $C_7^{eff}$, i.e., 
$C_7^{eff}=0.313$ the situation 
is similar to the previous case, but the magnitude of $\lla P_L^- + P_L^+ \rra$ 
is smaller. Figs. (3) and (4) are the same as Figs.(1) and (2) but for the
$B \rar K \tau^+ \tau^-$ decay. In this case the difference of the dependence 
of the longitudinal polarization $\lla P_L^- \rra$ on new Wilson coefficients
from the muon case is as follows: In the muon case
$\lla P_L^-\rra$ is negative for all values of the new Wilson coefficients
while for the tau case $\lla P_L^- \rra$ can receive both values, for
example for $C_T<1$, $\lla P_L^- \rra$ is positive, and for $C_T>1$, 
$\lla P_L^- \rra$ is negative.

It is obvious from Fig. (4) that if the values of the new Wilson coefficients
$C_{LRRL}$, $C_{LRLR}$, $C_{RLLR}$, $C_{RLRL}$ and $C_{TE}$ 
are negative (positive), $\lla P_L^- + P_L^+ \rra$ is negative (positive). 
Absolutely similar situation takes place 
for $C_7^{eff}>0$. 
For these reasons determination of the sign and of course magnitude of 
$\lla P_L^- + P_L^+\rra$ can give promising
information about new physics. 

In Figs. (5) and (6) the dependence of the average transversal polarization
$\lla P_T^- \rra$ and the combination $\lla P_T^- - P_T^+ \rra$ on the new
Wilson coefficients are presented for the $B \rar K \mu^+ \mu^-$ decay, 
respectively. 
We observe from Fig. (5) that the average
transversal polarization is strongly dependent only on $C_{LRRL}$
and $C_{RLRL}$ and quite weakly to remaining Wilson coefficients. It is also
interesting to note that for the negative (positive) values of these scalar
coefficients $\lla P_T^- \rra$ is negative (positive).
For the $\lla P_T^- - P_T^+ \rra$ case, there appears strong
dependence on all four scalar interactions with coefficients
$C_{LRRL},~C_{RLLR},~C_{LRLR},~C_{RLRL}$. The behavior of this combined
average transversal polarization 
is identical for the coefficients $C_{LRLR},~C_{RLLR}$ and 
$C_{LRRL},~C_{RLRL}$ in pairs, so that four lines responsible for these interactions
appear only to be two. Moreover $\lla P_T^- - P_T^+ \rra$ is negative
(positive) for the negative (positive) values of the new Wilson coefficients 
$C_{LRRL}$ and $C_{RLRL}$ and positive (negative) 
for the coefficients $C_{LRLR}$ and $C_{RLLR}$.   
Remembering that in SM, in massless lepton case $\lla P_T^- \rra \approx 0$ and 
$\lla P_T^- - P_T^+ \rra \approx 0$. Therefore  determination of the signs
and magnitudes of 
$\lla P_T^- \rra$ and $\lla P_T^- - P_T^+ \rra$ can give quite a useful
information about the existence of new physics. For the choice of
$C_7^{eff}=0.313$, apart from the minor differences in their magnitudes, the behaviors 
of $\lla P_T^- \rra$ and $\lla P_T^- - P_T^+ \rra$
are similar as in the previous case. 

As is obvious from Figs. (7) and (8), $\lla P_T^- \rra$ and 
$\lla P_T^- -  P_T^+\rra$ show stronger dependence only on
$C_T$ for the $B \rar K \tau^+ \tau^-$ decay. 
Again $\lla P_T^- \rra$ and  $\lla P_T^- - P_T^+ \rra$ 
change sign at $C_T \approx - 1$. 
As has already been noted, determination of the sign and
magnitude of  $\lla P_T^- \rra$ and $\lla P_T^- - P_T^+ \rra$     
are useful hints in looking for new physics.

Note that for simplicity all new Wilson coefficients in this work are 
assumed to be real. Under this condition 
$\lla P_N^- \rra$ and $\lla P_N^- + P_N^+ \rra$
have non-vanishing values coming from the imaginary part of SM, i.e., from
$C_9^{eff}$. From Figs. (9) and (10) we see that  $\lla P_N^- \rra$ 
and $\lla P_N^- + P_N^+ \rra$ are strongly
dependent on all scalar type interactions for the $B \rar K \mu^+ \mu^-$ decay. 
Similar behavior takes place for the $B \rar K \tau^+ \tau^-$ decay as well. 
The change in sign and 
magnitude of both $\lla P_N^- \rra$ and $\lla P_N^- + P_N^+ \rra$ that are 
observed in these figures is an indication of the fact that an experimental 
verification of them can give unambiguous information about new physics.

In the present work we analyze the possibility of pinning down new physics
beyond SM by studying lepton polarizations only. It follows from 
Eq. (\ref{unp}) that the branching ratio of the
$B \rar K \ell^+ \ell^-$ decay depends also on the new Wilson coefficients
and hence we expect that it can give information about new physics. In this
connection there follows the question: Can one establish new physics by
studying the lepton polarizations only? In other words, are there regions of
the new Wilson coefficients $C_X$ in which the value of the branching ratio
coincides with that of the SM prediction, but the lepton polarizations would
not? In order to answer this question, we present in Figs. (11)--(14) the
dependence of the branching ratio on the average and combined average
polarizations of the leptons. In these figures the value of the branching
ratio ranges between the values $10^{-7} \le {\cal B} (B \rar K \tau^+
\tau^-) \le 5 \times 10^{-7}$.
These figures depict that there indeed exist such regions of $C_X$ in which 
the value of the branching ratio does agree with the SM result, while the
lepton polarizations differ from the SM prediction. It follows from the pair
of Figs. (3), (11); (7), (13) and (8), (14), that if $C_T$ lies in the
region $-2 \le C_T \le 0$, the above--mentioned condition, i.e.,
mismatch of the polarizations in the standard model and the new physics, is
fulfilled. On the other hand one can immediately see from Fig. (12) that
such a region for the combined average longitudinal 
lepton polarization does not exist
and hence it is not suitable in search of new physics.
Note that in all figures intersection point of all curves correspond to the
SM case. This analysis allows us to conclude that there exists certain
regions of new Wilson coefficients for which study of the lepton
polarization itself can give promising information about new physics.
 
Finally, a few words about the detectibilty of the lepton
polarization asymmetries at $B$ factories or future hadron colliders, are in
order. As an estimation, we choose the averaged values of the longitudinal 
polarization of muon and transversal
and normal polarizations of the $\tau$ lepton, which are approximately close 
to the SM prediction, i.e., $\lla P_L \rra \simeq -0.9$, 
$\lla P_T \rra \simeq 0.6$ and $\lla P_N \rra \simeq -0.01$.
Experimentally, to measure an
asymmetry $\lla P_i \rra$ of a decay with the branching ratio $B$ at the 
$n \sigma$ level, the required number of events is given by the formula  
$N = n^2 / ({\cal B} \lla P_i
\rra^2$. It follows from this expression that to observe the lepton
polarizations $\lla P_L \rra$, $\lla P_T \rra$ and $\lla P_N \rra$ in 
$B \rar K \tau^+ \tau^-$ decay at $1\sigma$ level, the expected number
of events are $N=(1;3;10^4)\times10^7$, respectively.
On the other hand, The number of $B \bar B$ pairs that is expected to be 
produced at $B$ factories is about $N \sim 5\times 10^8$. A
comparison of these numbers allows us to conclude that while
measurement of the normal polarization of the $\tau$ lepton is 
impossible, measurements of the longitudinal polarization of muon
and transversal polarization of $\tau$ lepton 
could be accessible at $B$ factories.

\newpage
\section*{Figure captions}
{\bf Fig. (1)} The dependence of the average longitudinal polarization asymmetry
$\lla P_L^- \rra$ of muon on the new Wilson coefficients. \\ \\
{\bf Fig. (2)} The dependence of the combined average longitudinal
polarization asymmetry $\lla P_L^- + P_L^+\rra$ of $\ell^-$ and $\ell^+$ 
on the new Wilson coefficients for the $B \rar K \mu^- \mu^+$ decay. \\ \\ 
{\bf Fig. (3)} The same as in Fig. (1), but for the 
$B \rar K \tau^- \tau^+$ decay. \\ \\
{\bf Fig. (4)} The same as in Fig. (2), but for the 
$B \rar K \tau^- \tau^+$ decay. \\ \\
{\bf Fig. (5)} The same as in Fig. (1), but for the average transversal
polarization asymmetry $\lla P_T^- \rra$ of muon. \\ \\
{\bf Fig. (6)} The same as in Fig. (2), but for the transversal polarization
asymmetry $\lla P_T^- - P_T^+\rra$. \\ \\
{\bf Fig. (7)} The same as in Fig. (5), but for the 
$B \rar K \tau^- \tau^+$ decay. \\ \\
{\bf Fig. (8)} The same as in Fig. (6), but for the 
$B \rar K \tau^- \tau^+$ decay. \\ \\
{\bf Fig. (9)} The dependence of the average normal asymmetry
$\lla P_N^- \rra$ of muon on the new Wilson coefficients. \\ \\
{\bf Fig. (10)} The dependence of the combined average normal polarization 
asymmetry $\lla P_N^- + P_N^+\rra$ on the new Wilson coefficients
for the $B \rar K \mu^- \mu^+$ decay. \\ \\
{\bf Fig. (11)} Parametric plot of the correlation between the integrated
branching ratio ${\cal B}$ (in units of $10^{-7}$) and the  
average longitudinal lepton
polarization asymmetry $\lla P_L^-\rra$
as function of the new Wilson coefficients as indicated
in the figure, for the $B \rar K \tau^- \tau^+$ decay. \\ \\
{\bf Fig. (12)} The same as in the Fig. (11), but for the combined
average longitudinal lepton polarization asymmetry $\lla P_L^- + P_L^+
\rra$.\\ \\                                               
{\bf Fig. (13)} The same as in Fig. (11), but for the 
average transversal lepton polarization asymmetry $\lla P_T^-\rra$ . \\ \\
{\bf Fig. (14)} The same as in the Fig. (13), but for the combined
average transversal lepton polarization asymmetry $\lla P_T^- - P_T^+
\rra$.\\ \\

\newpage

\begin{figure}
\vskip 1cm
    \includegraphics{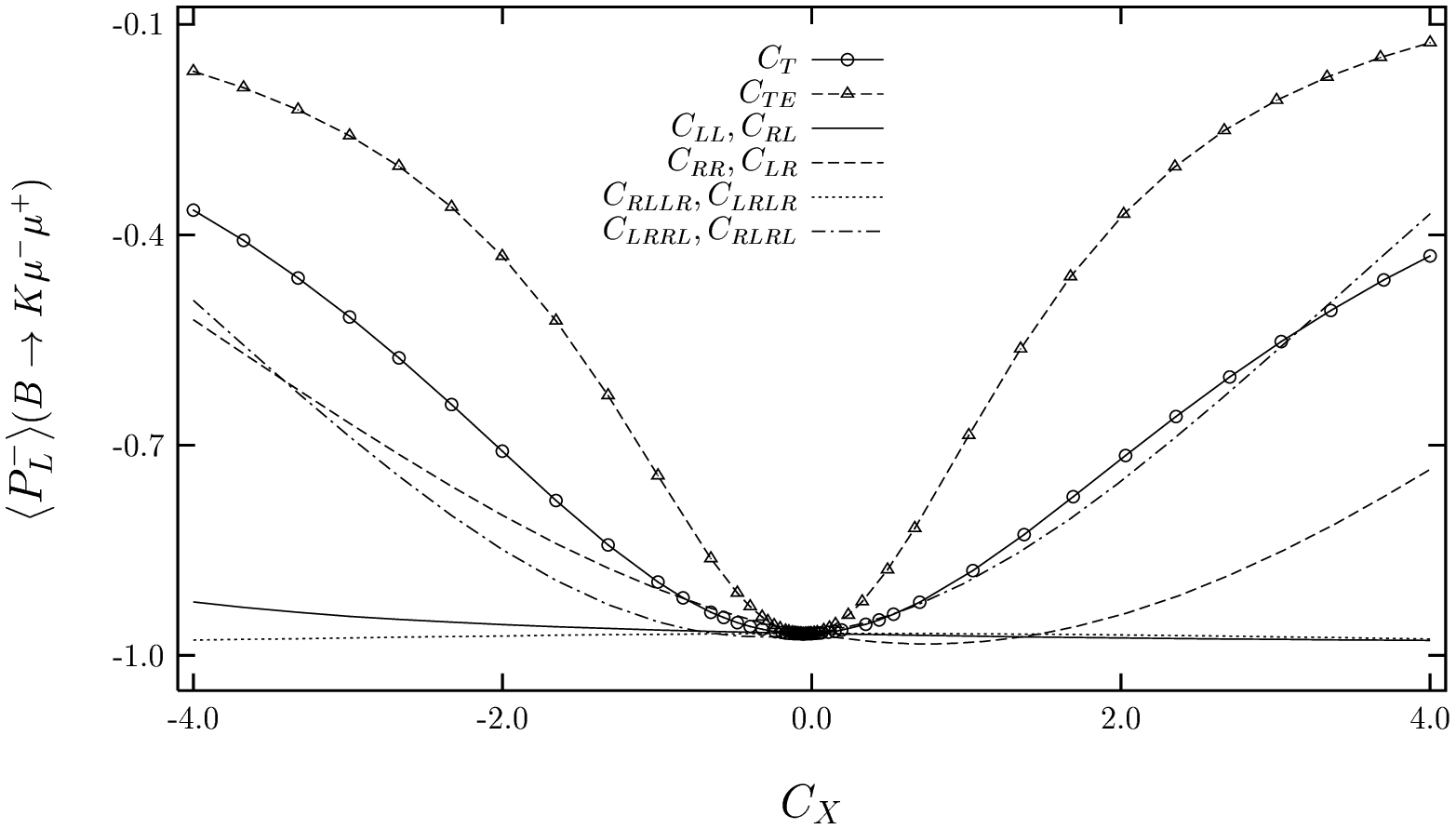}
\vskip 8.1cm
\caption{}
\end{figure}

\begin{figure}
\vskip 1.5 cm
    \includegraphics{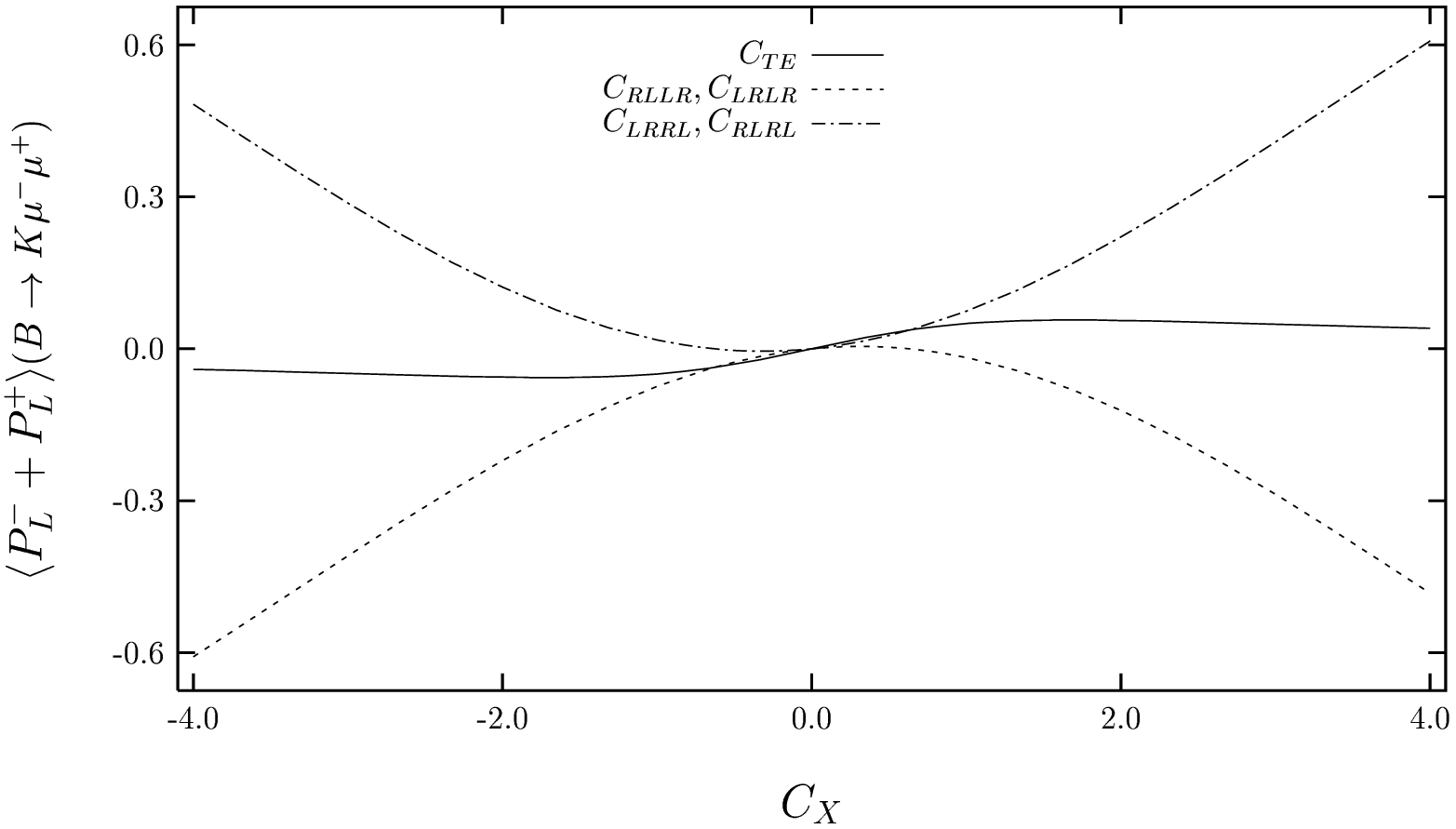}
\vskip 9. cm
\caption{}
\end{figure}

\begin{figure}
\vskip 1.5 cm
    \includegraphics{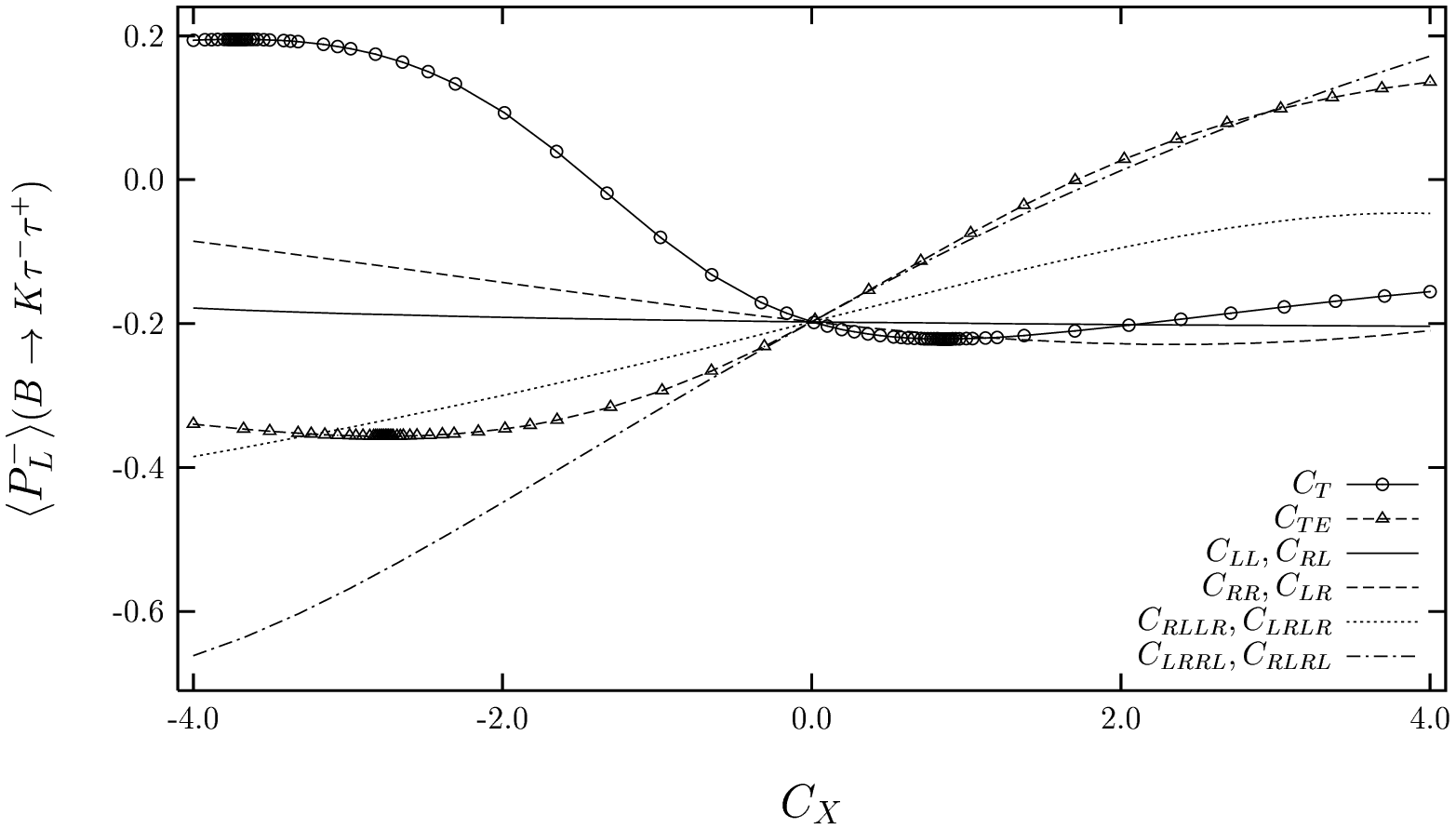}
\vskip 9. cm
\caption{}
\end{figure}

\begin{figure}
\vskip 1cm
    \includegraphics{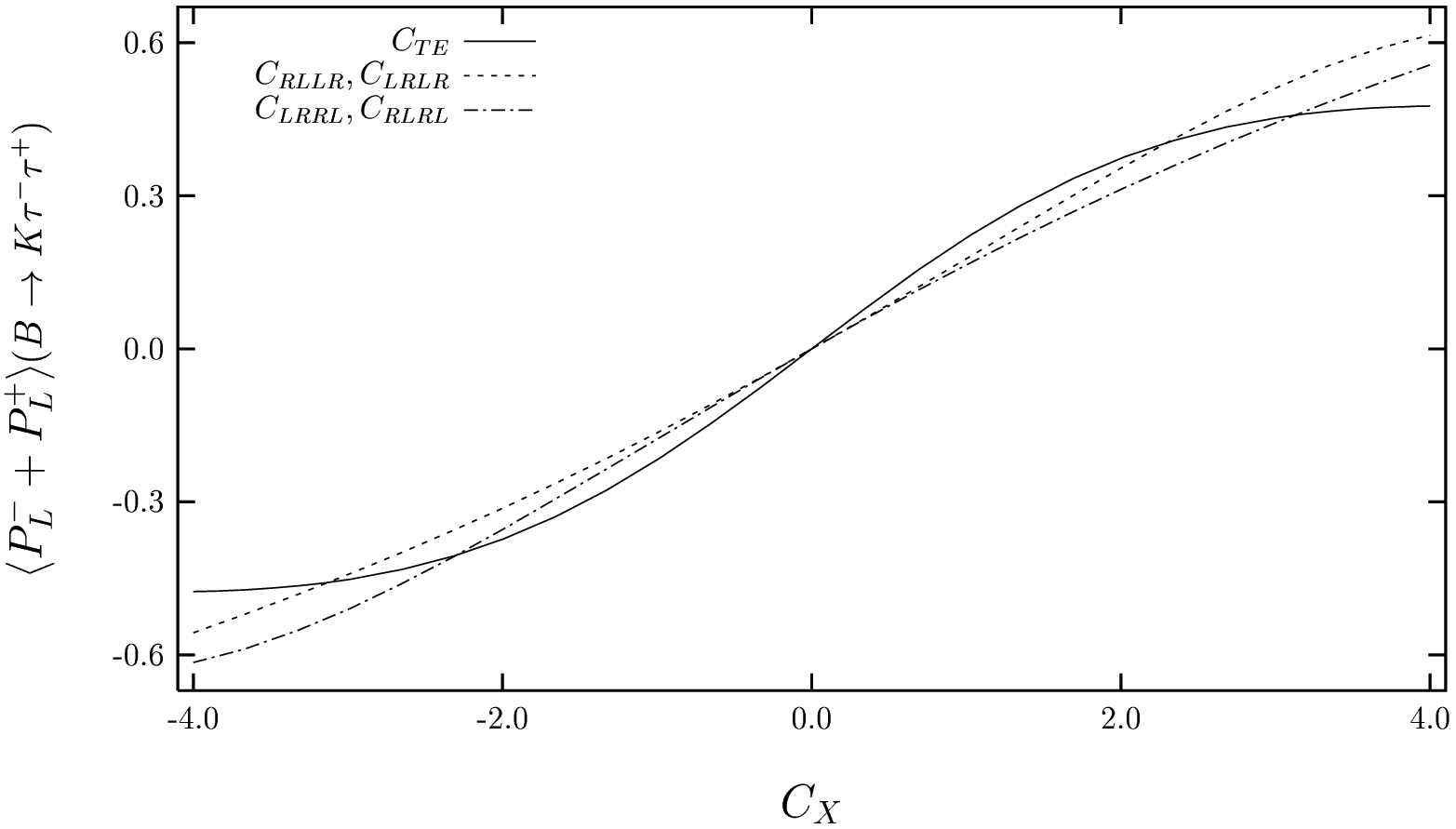}
\vskip 8.1cm
\caption{}
\end{figure}

\begin{figure}
\vskip 1.5 cm
    \includegraphics{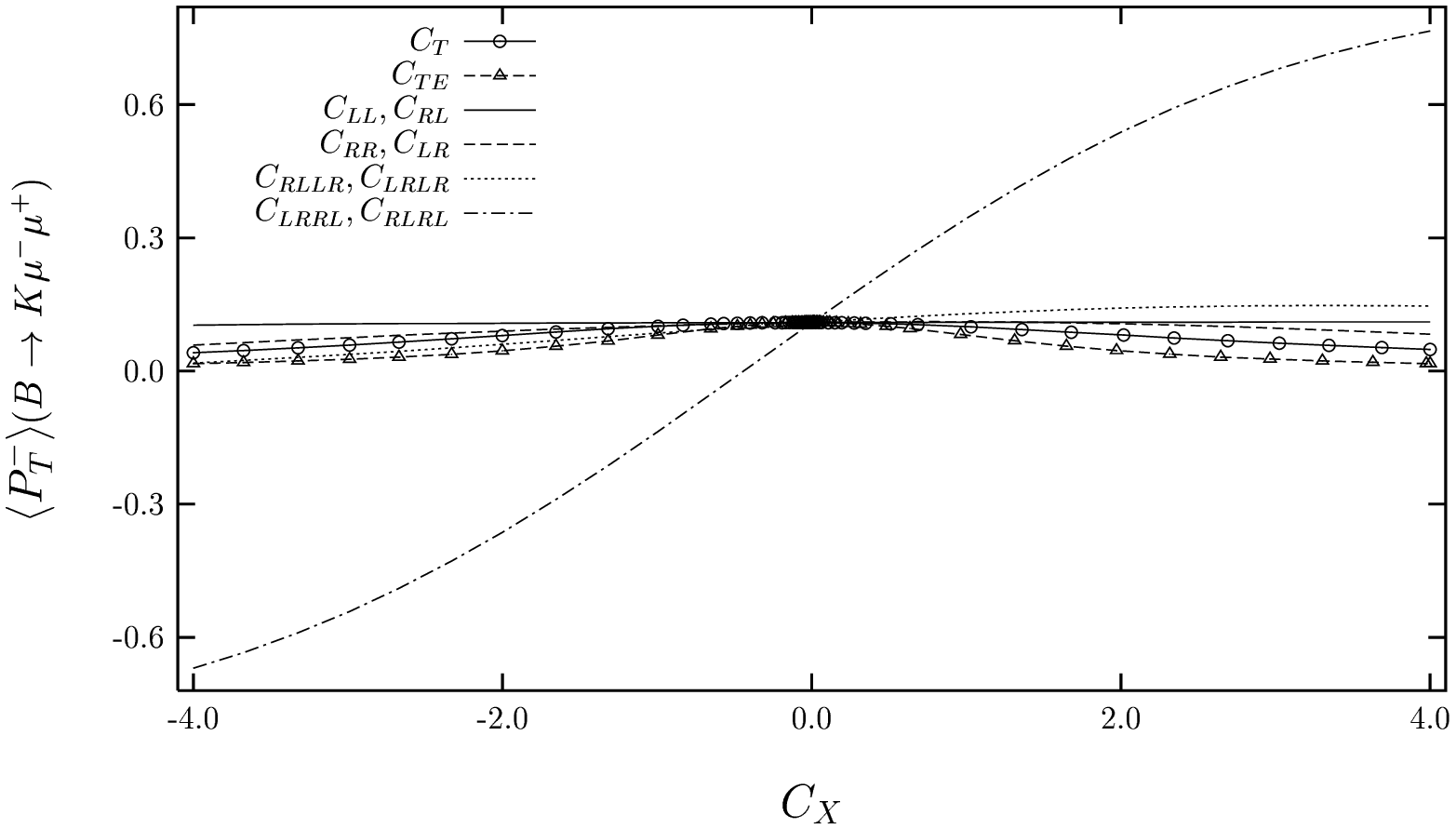}
\vskip 9. cm
\caption{}
\end{figure}

\begin{figure}
\vskip 1cm
    \includegraphics{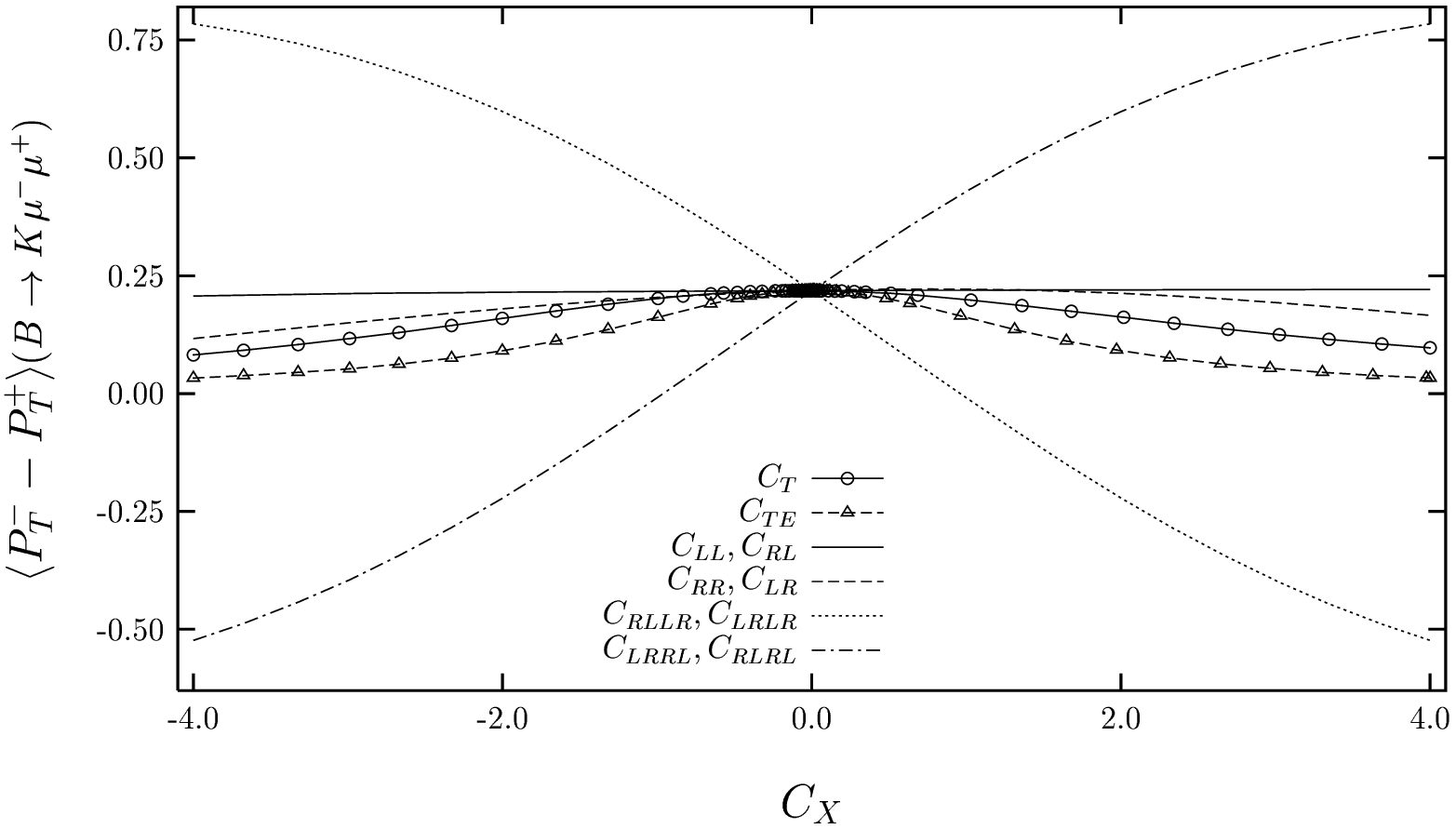}
\vskip 8.1cm
\caption{}
\end{figure}

\begin{figure}
\vskip 1.5 cm
    \includegraphics{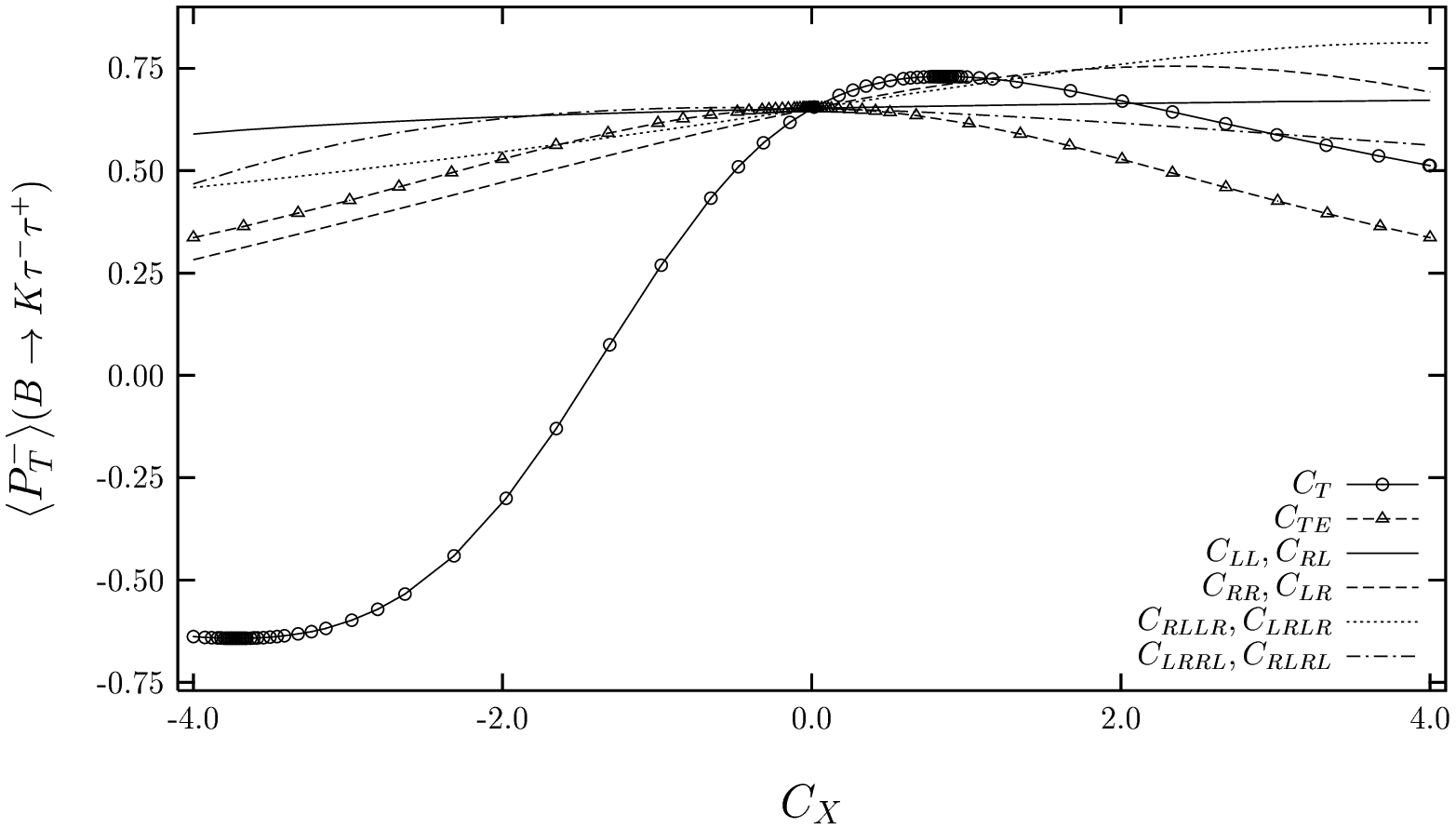}
\vskip 9. cm
\caption{}
\end{figure}

\begin{figure}
\vskip 1cm
    \includegraphics{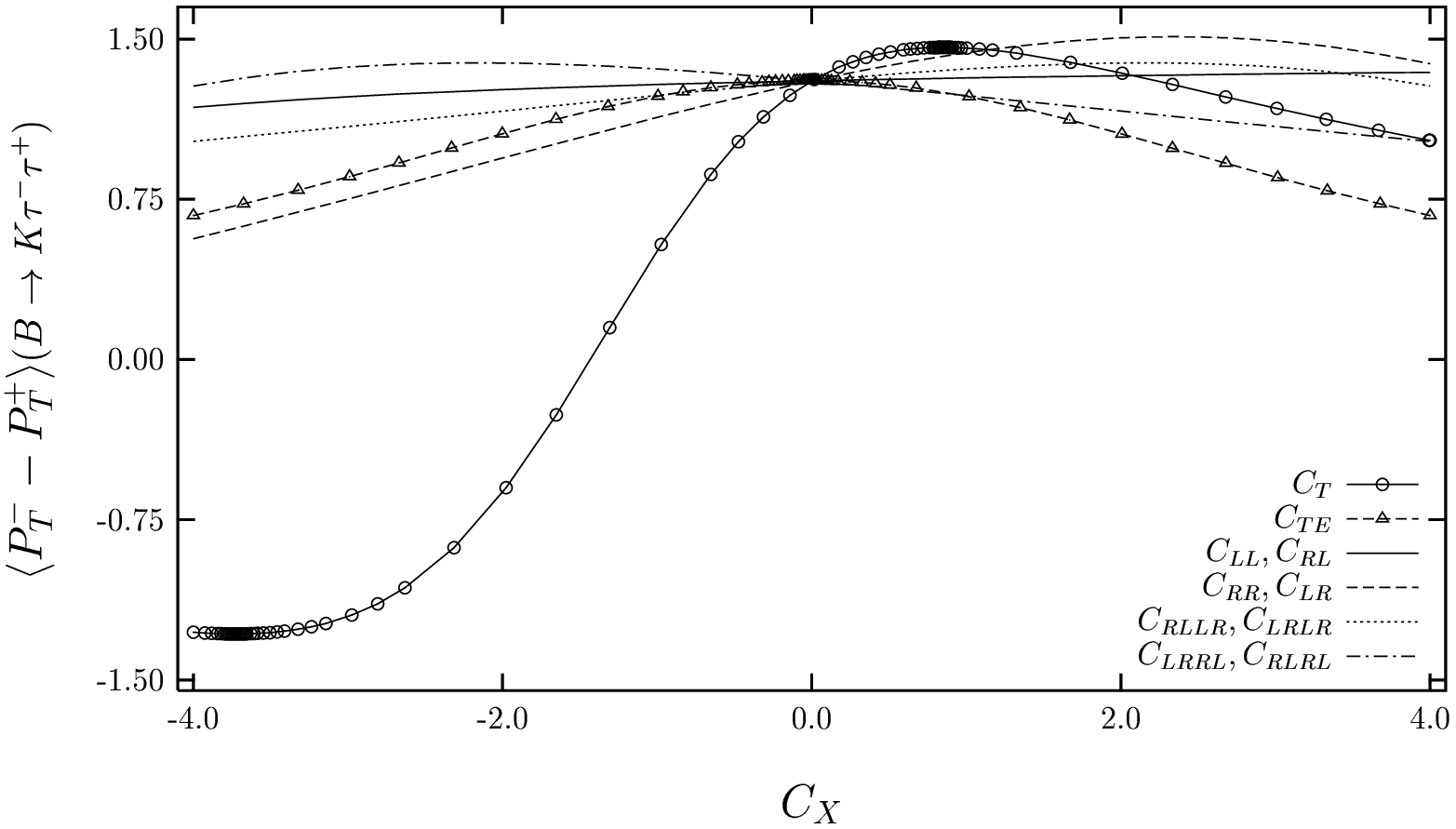}
\vskip 8.1cm
\caption{}
\end{figure}

\begin{figure}
\vskip 1.5 cm
    \includegraphics{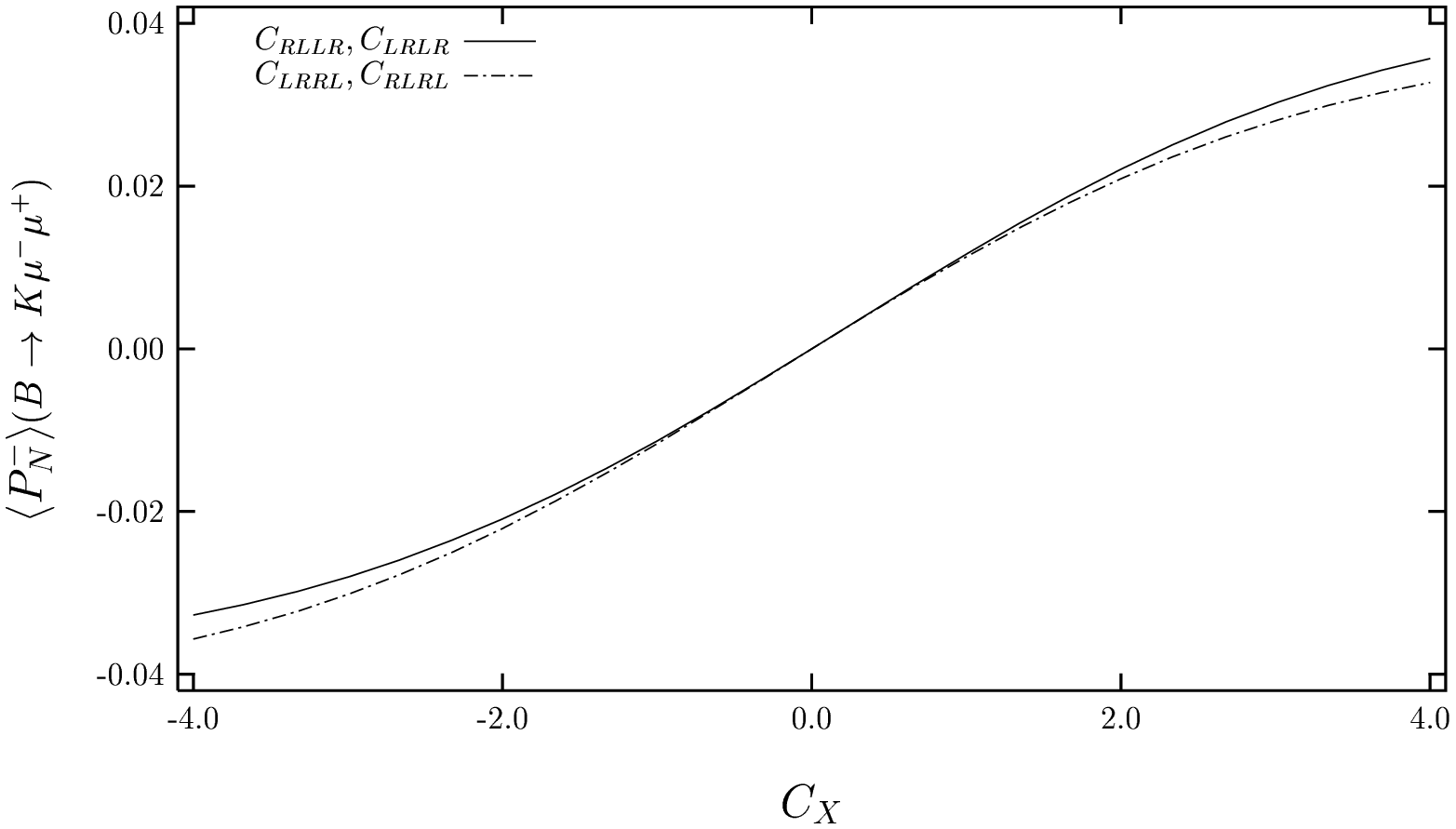}
\vskip 9. cm
\caption{}
\end{figure}

\begin{figure}
\vskip 1cm
    \includegraphics{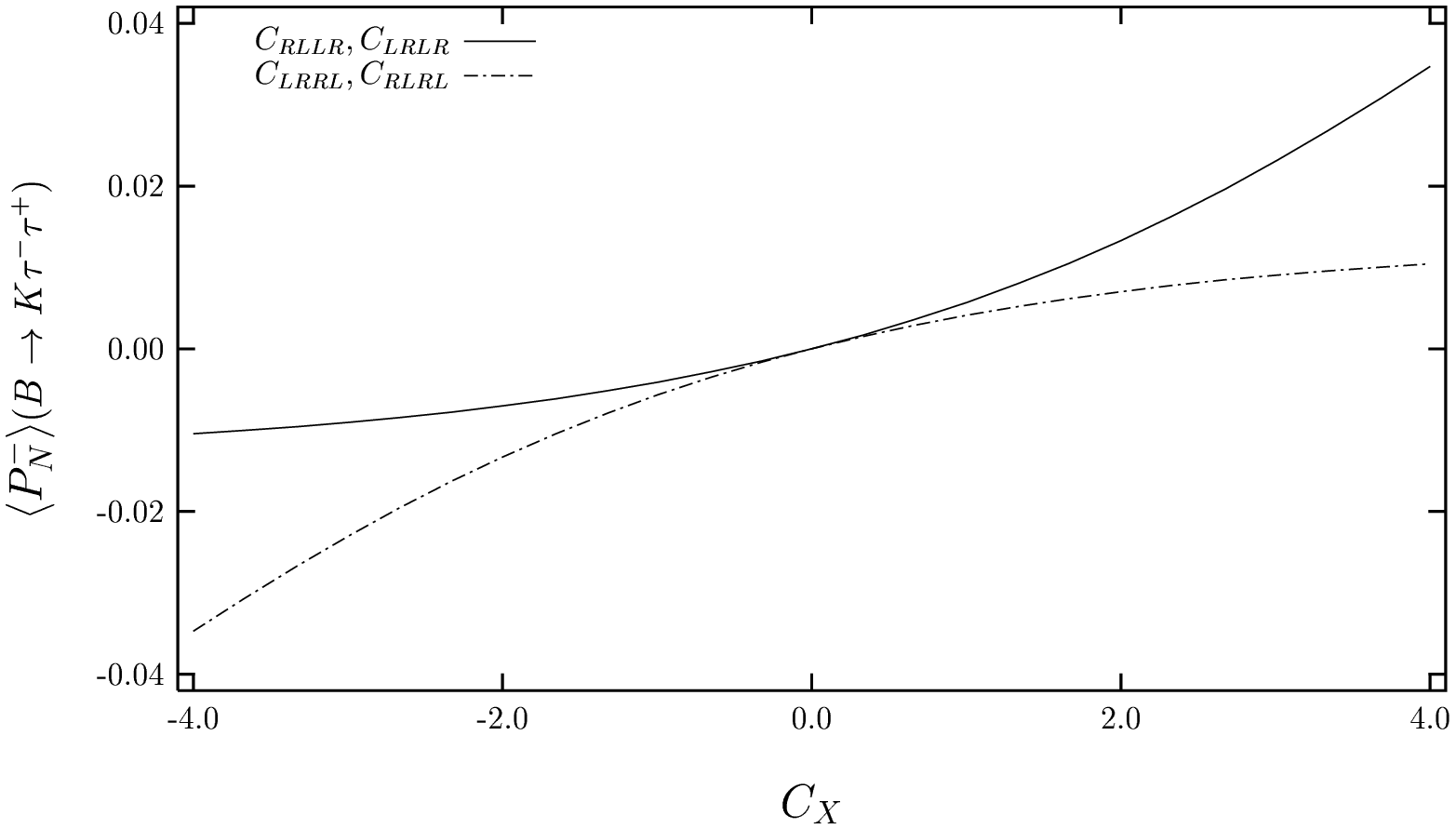}
\vskip 8.1cm
\caption{}
\end{figure}

\begin{figure}
\vskip 1.5 cm
    \includegraphics{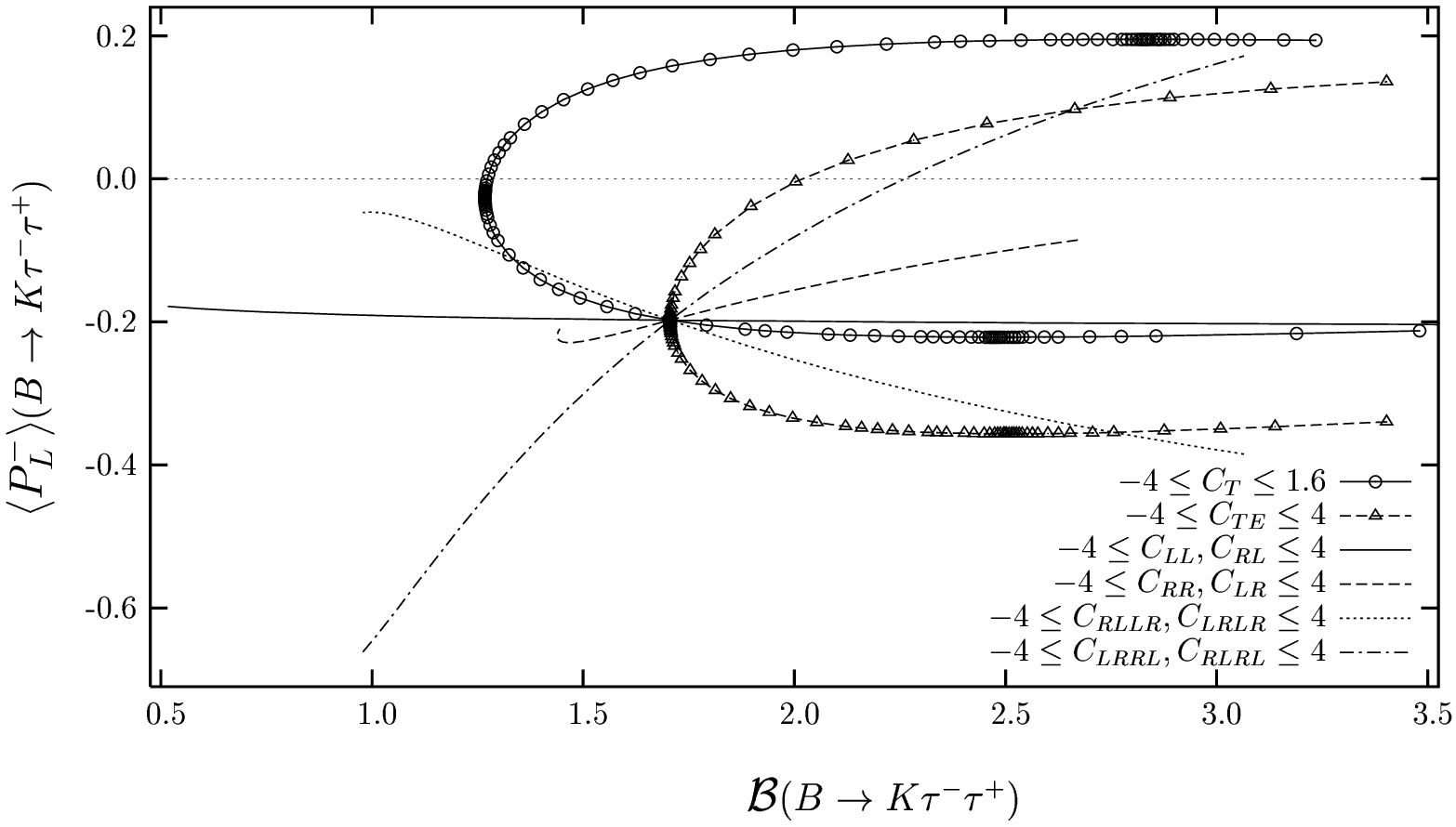}
\vskip 9. cm
\caption{}
\end{figure}

\begin{figure}
\vskip 1.5 cm
    \includegraphics{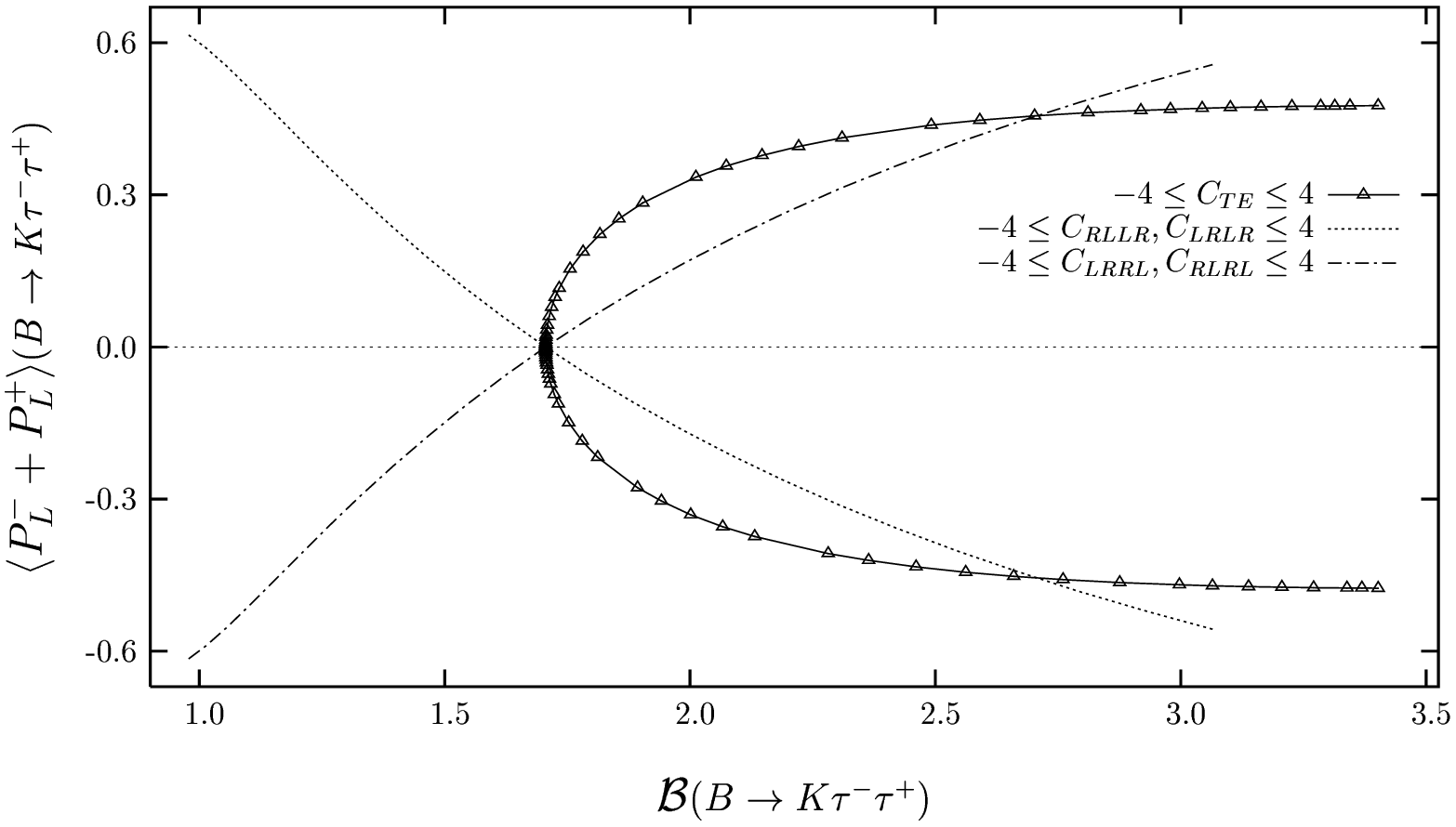}
\vskip 9. cm
\caption{}
\end{figure}

\begin{figure}
\vskip 1.5 cm
    \includegraphics{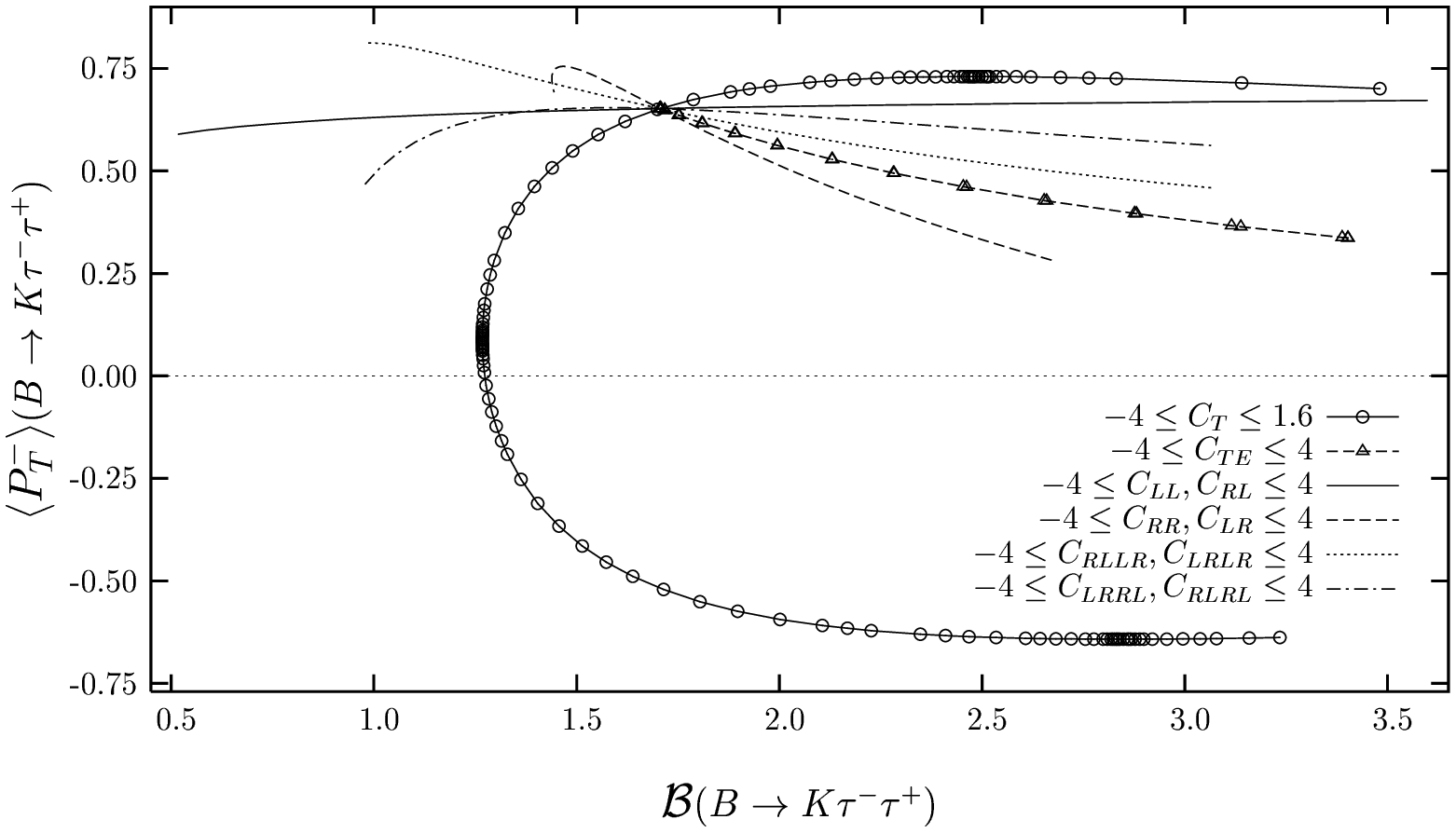}
\vskip 9. cm
\caption{}
\end{figure}

\begin{figure}
\vskip 1.5 cm
    \includegraphics{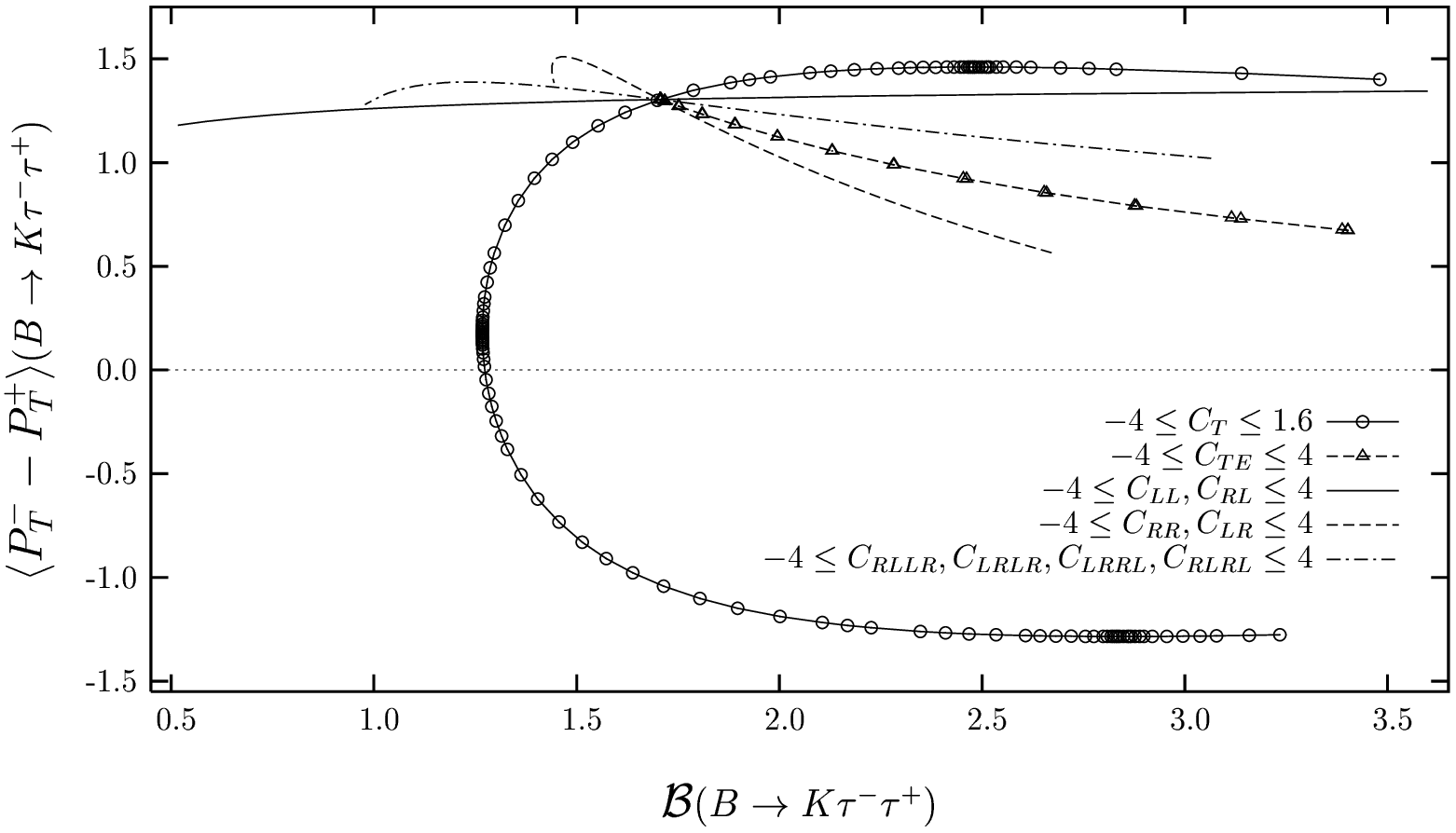}
\vskip 9. cm
\caption{}
\end{figure}

\end{document}